\documentstyle[12pt]{article}
\begin{document}

\author{Mariam Saleh Khan, M. Jamil Aslam, \and Amjad Hussain Shah Gilani, and
Riazuddin \\
National Center for Physics, Quaid-i-Azam University,\\
Islamabad 45320, Pakistan}
\title{Form factors and branching ratio for the $B\rightarrow l\nu \gamma $ decay }
\date{The Date }
\maketitle

\begin{abstract}
Form factors parameterizing radiative leptonic decays of heavy mesons $%
\left( B^{+}\rightarrow \gamma l^{+}\nu _{l}\right) $ for photon energy are
computed in the language of dispersion relation. The contributing states to
the absorptive part in the dispersion relation are the multiparticle
continum, estimated by quark triangle graph and resonances with quantum
numbers $1^{-}$ and $1^{+}$ which includes $B^{*}$ and $B_{A}^{*}$ and thier
radial excitations, which model the higher state contributions. Constraints
provided by the asymptotic behavior of the structure dependent amplitude,
Ward Identities and gauge invariance are used to provide useful information
for parameters needed. The couplings $g_{BB^{*}\gamma }$ and $%
f_{BB_{A}^{*}\gamma }$ are predicted if we restrict to first radial
excitation; otherwise using these as an input the radiative decay coupling
constants for radial excitations are predicted. The value of the branching
ratio for the process $B^{+}\rightarrow \gamma \mu ^{+}\nu _{\mu }$ is found
to be in the range $0.5\times 10^{-6}$. A detailed comparison is given with
other approaches.
\end{abstract}

\section{Introduction}

In spite of small branching ratio, the radiative $B$-meson decay $%
(B\rightarrow l\nu \gamma )$ is of viable interest because it contains
important information about weak and hadronic interactions of $B$-meson.
Furthermore, with the introduction of $B$-factories LHCb, BaBar, Belle and
CLEOb, the radiative $B$-meson decay can be studied with enough statistics.
Preliminary data from the CLEO collaboration indicates the limit on the
branching ratio ${\cal B}(B\rightarrow l\nu \gamma )$ which is: 
\begin{eqnarray*}
{\cal B}(B &\rightarrow &e\nu _{e}\gamma )<2.0\times 10^{-4} \\
{\cal B}(B &\rightarrow &\mu \nu _{\mu }\gamma )<5.2\times 10^{-5}
\end{eqnarray*}
at 90\% confidence level \cite{PRD56-11}. With the better statistics
expected from the upcoming $B$ factories, the observation and experimental
study of this decay could become soon feasible. It is therefore of some
interest to have a good theoretical control over the theoretical
uncertainties affecting the relevant matrix elements.

The radiative leptonic decay $B^{+}\rightarrow l^{+}\nu _{l}\gamma $ has
received a great deal of attention in the literature \cite
{PRD51-111,PLB358-329, NPB649-349, NPB650-356, PLB372-331, PLB386-328,
MPLA11-1061, JHEP042003029, PLB361-137, PRD57-5697, PRD61-114510,
PRD64-097503} as a means of probing aspects of the strong and weak
interactions of a heavy quark system. The presence of the additional photon
in the final state can compensate for the helicity suppression of the decay
rate present in purely leptonic mode. As a result, the branching ratio for
the radiative leptonic mode can be as large as $10^{-6}$ for the $\mu ^{+}$
case \cite{PLB361-137}, which would open up a possibility for directly
measuring the decay constant $f_{B}$ \cite{PLB386-328}. A study of this
decay can offer also useful information about the Cabibbo-Kobayashi-Maskawa
(CKM) matrix element $\left| V_{ub}\right| $ \cite{PRL10-531,PTP49-652}.

In the radiative $B$-decay process, there are two contributions to the
amplitude:

\begin{enumerate}
\item  inner bremsstrahlung (IB) and

\item  the structure dependent (SD) contribution which depends on the vector
and axial vector form factor $F_{V}$ and $F_{A}$ respectively.
\end{enumerate}

The IB contribution to the decay amplitude is associated with the tree
diagrams shown in Figs. 1a and 1b, and SD contribution is associated with
Fig. 1c.

In this paper, we will study the radiative leptonic $B$ decays of $%
B^{+}\rightarrow l^{+}\nu _l\gamma $. The IB part is still helicity
suppressed \cite{PRD51-111}, while the SD one is free of the suppression 
\cite{PRD15-709}. Therefore, the radiative decay rates of $B^{+}\rightarrow
l^{+}\nu _l\gamma $ $\left( l=e,\mu \right) $ could have an enhancement with
respect to the purely leptonic modes of $B^{+}\rightarrow l^{+}\nu _l$ due
to the SD contributions in spite of the electromagnetic coupling constant $%
\alpha $. With the possible large branching ratios, the radiative leptonic $%
B $ decays could be measured in the future experiments at hadronic
colliders, such as BTeV and CERN Large Hadron Collider (LHC-B) experiments 
\cite{SS9709500}.

The paper is organized as follows. In Sec. 2, we present the decay
kinematics and current matrix elements for $B^{+}\rightarrow l^{+}\nu
_{l}\gamma $. In section 3, we discuss the various contributions to the
absorptive part of the SD amplitude $iH_{\mu \nu }$, needed in the
dispersion relation. This include multiparticle continuum and resonances
with quantum numbers $1^{-}$ and $1^{+}$. The resonances include $B^{*}$ and 
$B_{A}^{*}$ mesons and their radial excitations, which model the higher
states. The continuum is estimated by quark triangle graphs. In Sec. 4, the
asymptotic behavior of the SD amplitude is studied. This provides a usual
constraint on the residues of the resonance contribution, in terms of the
continuum contribution. In Sec. 5, we discuss Ward Identities which together
with gauge invariance relates various form factors. These identities which
are expected to hold below the resonance regime, fix the normalization of
the forms at $q^{2}=0$ in terms of universal function $g_{+}\left( 0\right) $
as well as another constraint on the residues. Thus in our approach, a
parametrization of $q^{2}$ dependence of form factors is not approximated by
single pole contribution. But this parametrization is dictated by
considerations mentioned above and also predict the coupling constants of $%
1^{-}$ and $1^{+}$ resonances with photon if we restrict to one radial
excitation; otherwise using these as input, the radiative coupling constants
of radial excitations are predicted. In this and other aspects our approach
is different from the others mentioned previously. Our approach is closest
to the one used in \cite{Burdman2} for $B\rightarrow \pi l\nu _{l}$. We
calculate the decay branching ratios in Sec. 5. We give our conclusions in
Sec. 6.

\section{Decay kinematics and current matrix elements}

We consider the decay 
\begin{equation}
B^{+}(p)\rightarrow l^{+}(p_{l})\nu _{l}(p_{\nu })\gamma (k),  \label{1}
\end{equation}
where $l$ stands for $e$ or $\mu $, and $\gamma $ is a real photon with $%
k^{2}=0$. The decay amplitude for radiative leptonic decay of $%
B^{+}\rightarrow l^{+}\nu _{l}\gamma $ can be written in two parts, $M_{IB}$
and $M_{SD}$, as follows: 
\begin{equation}
M(B^{+}\rightarrow l^{+}\nu _{l}\gamma )=M_{IB}+M_{SD}  \label{2}
\end{equation}
in terms of two emission types of real photon from $B^{+}\rightarrow
l^{+}\nu _{l}$. They are given by \cite
{PRB136-1160,PRs88-151,SJPN7-286,NPB396-81} 
\begin{eqnarray}
M_{IB} &=&ie\frac{G_{F}}{\sqrt{2}}V_{ub}f_{B}m_{l}\epsilon _{\mu }^{*}L^{\mu
}  \label{3} \\
M_{SD} &=&-i\frac{G_{F}}{\sqrt{2}}V_{ub}f_{B}m_{l}\epsilon _{\mu }^{*}\tilde{%
H}^{\mu \nu }l_{\nu }  \label{4}
\end{eqnarray}
with 
\begin{eqnarray}
L^{\mu } &=&m_{l}\bar{u}(p_{\nu })\left( 1+\gamma _{5}\right) \left( \frac{
2p^{\mu }}{2p\cdot k}-\frac{2p_{l}^{\mu }+\not{k}\gamma ^{\mu }}{2p_{l}\cdot
k}\right) v(p_{l},s_{l}),  \label{5} \\
l^{\mu } &=&\bar{u}(p_{\nu })\gamma ^{\mu }\left( 1+\gamma _{5}\right)
v(p_{l},s_{l}),  \label{6} \\
\tilde{H}^{\mu \nu } &=&iF_{V}(q^{2})\epsilon ^{\mu \nu \alpha \beta
}k_{\alpha }p_{\beta }-F_{A}(q^{2})\left( p\cdot kg^{\mu \nu }-p^{\mu
}k^{\nu }\right) ,  \label{7} \\
q^{\mu } &=&\left( p-k\right) ^{\mu }=\left( p_{l}+p_{\nu }\right) ^{\mu }.
\label{8}
\end{eqnarray}
Here $\epsilon _{\mu }^{*}$ denotes the polarization vector of the photon
with $k^{\mu }\epsilon _{\mu }^{*}\left( k\right) =0$. $p$, $p_{l}$, $p_{\nu
}$, and $k$ are the four momenta of $B^{+}$, $l^{+}$, $\nu $, and $\gamma $,
respectively, $s_{l}$ is the polarization vector of the $l^{+}$, $f_{B}$ is
the $B$ meson decay constant, and $F_{A}$, $F_{V}$ stand for two Lorentz
invariant amplitudes (form factors).

The term proportional to $L^{\mu }$ in (\ref{5}) does not contain unknown
quantities--it is determined by the amplitude of the non-radiative decay $%
B^{+}\rightarrow l^{+}\nu _{l}$. This part of the amplitude is usually
referred as ``inner bremsstrahlung (IB) contribution'', whereas the term
proportional to $H^{\mu \nu }$ is called ``structure dependent (SD)
contribution''.

The form factor $F_{A}\left( F_{V}\right) $ is related to the matrix element
of the axial (vector) current. The factors $f_{B}$ and $F_{V,A}$ are defined
by 
\begin{eqnarray}
\left\langle 0\left| \bar{u}\gamma ^{\mu }\gamma _{5}b\right|
B(p)\right\rangle &=&-if_{B}p^{\mu }  \label{9} \\
\left\langle \gamma \left( k\right) \left| \bar{u}\gamma ^{\mu }\gamma
_{5}b\right| B(p)\right\rangle &=&-\left[ \left( \epsilon ^{*}\cdot p\right)
k^{\mu }-\epsilon ^{*\mu }\left( p\cdot k\right) \right] F_{A}(q^{2})
\label{10} \\
\left\langle \gamma \left( k\right) \left| \bar{u}\gamma ^{\mu }b\right|
B(p)\right\rangle &=&-i\epsilon ^{\mu \nu \alpha \beta }\epsilon _{\nu
}^{*}p_{\alpha }k_{\beta }F_{V}(q^{2})  \label{11}
\end{eqnarray}

In our phase convention, the form factors $F_{A}$ and $F_{V}$ are real in
the physical region 
\[
m_{l}^{2}\ll q^{2}\ll M_{B}^{2} 
\]
where $q$ is the momentum transfer. The kinematics of the decay needs two
variables, for which we choose the conventional quantities and in the rest
frame of $B$%
\begin{eqnarray}
x &=&\frac{2p\cdot k}{M_{B}^{2}}=\frac{2E_{\gamma }}{M_{B}}  \label{12} \\
y &=&\frac{2p\cdot p_{l}}{M_{B}^{2}}=\frac{2E_{l}}{M_{B}}  \label{13}
\end{eqnarray}
and the angle $\theta _{l\gamma }$ between the photon and the charged lepton
is related to $x$ and $y$ by 
\begin{equation}
x=\frac{1}{2}\frac{\left( 2-y+\sqrt{y^{2}-4r_{l}}\right) \left( 2-y-\sqrt{%
y^{2}-4r_{l}}\right) }{2-y+\sqrt{y^{2}-4r_{l}}\cos \theta _{l\gamma }}.
\label{14}
\end{equation}
In terms of these quantities, one can write the momentum transfer as 
\begin{equation}
q^{2}=M_{B}^{2}\left( 1-x\right) ,\,\,\,\,\,\left( k^{2}=0\right) .
\label{15}
\end{equation}
We write the physical region of $x$ and $y$ as 
\begin{eqnarray}
0 &\leq &x\leq 1-r_{l},  \label{16} \\
1-x+\frac{r_{l}}{1-x} &\leq &y\leq 1+r_{l},  \label{17}
\end{eqnarray}
where 
\begin{equation}
r_{l}=\frac{m_{l}^{2}}{M_{B}^{2}}=\left\{ 
\begin{array}{ccc}
9.329\times 10^{-9} &  & \left( l=e\right) , \\ 
4.005\times 10^{-4} &  & \left( l=\mu \right) .
\end{array}
\right.  \label{18}
\end{equation}

\section{Dispersion Relations}

The structure dependent part, $H^{\mu \nu }$ is given by 
\begin{equation}
iH^{\mu \nu }=i\int d^{4}xe^{ik\cdot x}\left\langle 0\left| T\left(
j_{em}^{\mu }(x)J_{2}^{\nu }(0)\right) \right| B(p)\right\rangle  \label{19}
\end{equation}
We note that \cite{23} 
\begin{equation}
ik_{\mu }H^{\mu \nu }=if_{B}p_{\nu }{ ,}  \label{20}
\end{equation}
so that for the real photon we can write 
\begin{equation}
H^{\mu \nu }=\tilde{H}^{\mu \nu }+f_{B}\frac{p^{\mu }p^{\nu }}{p\cdot k}
\label{21}
\end{equation}
where $k_{\mu }\tilde{H}^{\mu \nu }=0$ and $\tilde{H}^{\mu \nu }$ is
parametrized as in Eq.(\ref{7}). The second term in (\ref{21}) is absorbed
in $M_{IB}$. The absorptive part is 
\begin{eqnarray}
Abs\left[ iH^{\mu \nu }\right] &=&\frac{1}{2}\int d^{4}xe^{ik\cdot
x}\left\langle 0\left| \left[ j_{em}^{\mu }(x),J_{2}^{\nu }(0)\right]
\right| B(p)\right\rangle  \nonumber \\
&=&\frac{1}{2}(2\pi )^{4}{[\sum_n }\left\langle 0\left|
j_{em}^{\mu }(0)\right| n\right\rangle \left\langle n\left| J_{2}^{\nu
}(0)\right| B(p)\right\rangle \delta ^{4}(k-p_{n})  \nonumber \\
&&\left. -{\sum_n }\left\langle 0\left| J_{2}^{\nu }(0)\right|
n\right\rangle \left\langle n\left| j_{em}^{\mu }(0)\right|
B(p)\right\rangle \delta ^{4}(k+p_{n}-p)]\right.  \label{22}
\end{eqnarray}
The $\delta $-function in the first term implies $p_{n}^{2}=k^{2}=0$ and
since there is no real particle with zero mass, the first term does not
contribute. Thus contributing to the absorptive part are all possible
intermediate states that couple to $B\gamma $ and annihilated by the weak
vertex $\left\langle 0\left| J_{2}^{\nu }(0)\right| n\right\rangle $. These
include the multiparticle continuum as well as resonances with quantum
numbers $1^{-}$ and $1^{+}$. Thus $\left( t=q^{2}\right) $%
\begin{eqnarray}
F_{V}(t) &=&\frac{g_{BB^{*}\gamma }}{M_{B^{*}}^{2}-t}f_{B^{*}}+\cdots 
\nonumber \\
&&  \label{23} \\
F_{A}(t) &=&\frac{f_{B_{A}^{*}B\gamma }}{M_{B_{A}^{*}}^{2}-t}%
f_{B_{A}^{*}}+\cdots  \nonumber
\end{eqnarray}
The ellipses stand for contributions from higher states with the same
quantum numbers. The couplings $g_{BB^{*}\gamma }$ and $f_{B_{A}^{*}B\gamma
} $ are defined as 
\begin{eqnarray}
\left\langle B^{*-}(q,\eta )\gamma \left( k,\epsilon \right) \mid
B^{-}\left( P\right) \right\rangle &=&ig_{B^{*}B\gamma }\varepsilon _{\alpha
\rho \mu \sigma }\epsilon ^{*\alpha }q^{\rho }\eta ^{*\mu }p^{\sigma } 
\nonumber \\
\left\langle B_{A}^{*-}(q,\eta )\gamma \left( k,\epsilon \right) \mid
B^{-}\left( P\right) \right\rangle &=&ig_{B_{A}^{*}B\gamma }(\epsilon
^{*}.\eta ^{*})-if_{B_{A}^{*}B\gamma }(q.\epsilon ^{*})(k.\eta ^{*}) 
\nonumber \\
\left\langle 0\left| i\bar{u}\gamma ^{\mu }b\right| B^{*}(q,\eta
)\right\rangle &=&f_{B^{*}}\eta ^{\mu }  \nonumber \\
\left\langle 0\left| i\bar{u}\gamma ^{\mu }\gamma _{5}b\right|
B_{A}^{*}(q,\eta )\right\rangle &=&f_{B_{A}^{*}}\eta ^{\mu }  \label{23a}
\end{eqnarray}
We assume that the contributions from the radial excitations of $B^{*}$ and $%
B_{A}^{*}$ dominate the higher state contribution. Thus we write 
\begin{eqnarray}
F_{V}(t) &=&\frac{R_{V}}{1-t/M_{B^{*}}^{2}}+\sum_{i}\frac{R_{V_{i}}}{%
1-t/M_{B_{i}^{*}}^{2}}+\frac{1}{\pi }\int_{S_{0}}^{M^{2}}\frac{{\Im}%
F_{V}^{{Cont}}(s)}{s-t-i\varepsilon }ds+\ldots  \nonumber \\
&&  \label{24} \\
F_{A}(t) &=&\frac{R_{A}}{1-t/M_{B_{A}^{*}}^{2}}+\sum_{i}\frac{R_{A_{i}}}{%
1-t/M_{B_{A_{i}}^{*}}^{2}}+\frac{1}{\pi }\int_{S_{0}}^{M^{2}}\frac{{\Im}%
F_{A}^{{Cont}}(s)}{s-t-i\varepsilon }ds+\ldots  \nonumber
\end{eqnarray}
where ellipses stands for the contributions from the region for much larger
than the physical mass of heavy resonances up to $\infty $. Here, $M$ is a
cut off near the first radial excitation of $M_{B^{*}}$ or $M_{B_{A}^{*}}$
and $S_{0}=M_{B}+m_{\pi }$, and 
\begin{eqnarray}
R_{V} &=&\frac{g_{BB^{*}\gamma }}{M_{B^{*}}^{2}}f_{B^{*}}  \nonumber \\
&&  \label{25} \\
R_{A} &=&\frac{f_{B_{A}^{*}B\gamma }}{M_{B_{A}^{*}}^{2}}f_{B_{A}^{*}} 
\nonumber
\end{eqnarray}
$R_{V_{i}}$ and $R_{A_{i}}$ are the corresponding quantities for the radial
excitations with masses $M_{B_{i}^{*}}$ and $M_{B_{A_{i}}^{*}}$. In the next
section we develop the constraints on some of the parameters appearing in
the above equations.

If we model the continuum contribution by quark triangular graph (similar
calculations exist in the literature \cite{IJMPA17-4927}), we obtain 
\begin{equation}
F_{V}^{{Cont}}=F_{A}^{{Cont}}=\frac{f_{B}}{M_{B}}\left\{ \frac{
Q_{u}}{\bar{\Lambda}}-\frac{Q_{b}}{M_{B}}(1+\frac{\bar{\Lambda}}{M_{B}}%
)\right\} \frac{1}{1-q^{2}/M_{B}^{2}}  \label{26}
\end{equation}
where 
\begin{equation}
\bar{\Lambda}=M_{B}-m_{b},  \label{26Lambda}
\end{equation}
together with the term 
\[
(Q_{u}-Q_{b})f_{B}\frac{p^{\mu }p^{\nu }}{k\cdot p}=f_{B}\frac{p^{\mu
}p^{\nu }}{k\cdot p} 
\]
which appears in Eq. (\ref{21}). As is well known (see for example Ref. \cite
{nisgur}), the pole at $q^{2}=M_{B}^{2}$ in Eq. (\ref{26}) arises due to $u$ 
$\left( \bar{b}\right) $quark propagator which form one leg of quark $\Delta 
$, the other legs are the part of $B$ meson wave function.

\section{Asymptotic Behavior}

To get constraints on the residues $R_{i}$, it is useful to study the
asymptotic behavior of form factors $F_{V}$ and $F_{A}$. It has been argued
that the behavior of form factor for very large values of $\left| t\right| $
can be estimated reliably in perturbative QCD processes [pQCD]\cite{Brodsky,
Henley, Burdman2}. For $t\ll 0$ and for $\left| t\right| $ much larger than
the physical mass of heavy resonances, pQCD should yield a very good
approximation to the form factors. First we note that by vector meson
dominance 
\begin{equation}
\left\langle \gamma \left( k,\varepsilon ^{*}\left( k\right) \right) \left| 
\bar{u}\gamma ^{\mu }\left( 1-\gamma _{5}\right) b\right| B\left( p\right)
\right\rangle \simeq Q_{u}\frac{f_{\rho }}{m_{\rho }}\left\langle \rho
\left( k,\varepsilon ^{*}\left( k\right) \right) \left| \bar{u}\gamma ^{\mu
}\left( 1-\gamma _{5}\right) b\right| B\left( p\right) \right\rangle ,
\label{26a}
\end{equation}
where $f_{\rho }$, having dimensions of mass, is defined as 
\begin{equation}
\left\langle 0\left| \bar{u}\gamma ^{\mu }u\right| \rho \left( k,\varepsilon
\left( k\right) \right) \right\rangle =\frac{f_{\rho }}{m_{\rho }}%
\varepsilon ^{\mu }  \label{26b}
\end{equation}
Then using the methods employed in \cite{Henley}, it is eaisy to calculate
[only the diagram where gluon is emitted by the light quark in $\left( b\bar{%
u}\right) $ bound state and absorbed by the heavy quark contributes and is
by itself gauge invariant] $F^{pQCD}$: 
\begin{eqnarray}
F_{V}^{pQCD} &=&F_{A}^{pQCD}  \nonumber \\
&=&\left( Q_{u}\frac{f_{\rho }}{m_{\rho }}\right) \frac{32\pi \alpha
_{s}\left( t\right) }{3}\left( f_{B}f_{\rho }\right) m_{B}\left( \frac{1}{%
\varepsilon }\ln \varepsilon \right) \frac{1}{t^{2}}  \label{26c}
\end{eqnarray}
Here 
\begin{equation}
\varepsilon \sim {\cal O}\left( \frac{\Lambda _{QCD}}{m_{B}}\right)
\label{26d}
\end{equation}
and is governed by the tail end of the $B$ meson wave function characterized
by $\varepsilon $.

Now the asymptotic behavior of Eq. (\ref{24}), is given by 
\begin{equation}
F\left( q^{2}\right) \rightarrow -\frac{1}{q^{2}}\left[
RM^{2}+\sum_{i}R_{i}M_{i}^{2}+\frac{1}{\pi }\int_{S_{0}}^{M^{2}}{\Im}F^{%
{Cont}}(s)ds\right] .  \label{26e}
\end{equation}
Since $F^{pQCD}\left( t\right) $ is a reliable approximation to the form
factor for $t\rightarrow -\infty $, and $\left( tF^{pQCD}\right) \rightarrow
0$ in this limit, it follows that 
\begin{equation}
RM^{2}+\sum_{i}R_{i}M_{i}^{2}+c\simeq 0,  \label{convergence-new}
\end{equation}
where we have defined 
\begin{equation}
c=\frac{1}{\pi }\int_{S_{0}}^{M^{2}}{\Im}F^{{Cont}}(s)ds.
\label{26e1}
\end{equation}
The convergence relation (\ref{convergence-new}) is a model-independent
result and constitutes a very binding constraint for model building. In
other words, the various contributions in Eq. (\ref{26e}) may be individully
much larger than the$\left( tF^{pQCD}\left( t\right) \right) $ due to $%
\alpha _{s}\left( t\right) /t$ suppression, but there must be large
cancellations among the non-perturbative contributions in (\ref{26e}). This
is in the spirit of ref. \cite{Burdman2}. We will explore the resonant
contribution (in our model) in order to understand the effect of Eq. (\ref
{convergence-new}) on the behavior of form factors in the physical region.
The imposition of this constraint will lead to a very distinct behavior of
the photon momentum distribution, independently of how many resonances we
choose to keep. As the radial excitations of $B^{*}$ become heavier, they
are less relevant to the form factors since the spacing between the
consecutive radial excitations are expected to become narrower and narrower 
\cite{Quigg}. Thus, heavier resonances contribute with a smaller value even
in the narrow width approximation. Furthermore, as finite widths are
considered, the contribution of heavier and thus broader excitations are
additionally suppressed. This shows that the truncation of the sum over
resonances is a reasonable approximation.

For the resonances stated above we will study a constrained dispersive model
where only the first two radial excitations are kept. This is mainly for the
reason mentioned above. On the other hand, the ``minimal'' choice of keeping
only one radial excitation will determine $R_{1}$ in terms of $R$. The other
necessary ingredient to specify the model is the knowledge of the spectrum
of radial excitations. These resonances [(2S) and (3S) excitations of $B^{*}$%
] have not yet been observed in the $B$ systems. We will then rely on
potential model calculations for their masses \cite{Quigg}. These models
have been very successful in predicting the masses of orbitally excited
states and as such we are confident that the position of the radial
excitations does not introduce a sizeable uncertanity. The resultant
spectrum explicitly shows that the spacing among 1S, 2S, 3S states are, to
leading order, independent of heavy quark mass and, therefore, constitutes
the property of the light degrees of freedom. The spectrum of radial
excitations is given in Table 1, where the subindices 1 and 2 correspond to
the 2S and 3S excitation of the $B^{*}$, etc. Thus the convergence condition
(\ref{convergence-new}) now reads 
\begin{equation}
RM^{2}+R_{1}M_{1}^{2}+R_{2}M_{2}^{2}+c=0,  \label{convergence2}
\end{equation}

This condition leaves two free parameters $R_{1}$ and $R_{2}$ in the model.
This results in the correct scaling of form factors with the heavy meson
mass. Solving Eq. (\ref{convergence2}) for $R_{2}$ and using in Eq. (\ref
{26e}), we obtain 
\begin{equation}
F\left( q^{2}\right) =\frac{RM^{2}\left( M_{2}^{2}-M^{2}\right) }{\left(
M^{2}-q^{2}\right) \left( M_{2}^{2}-q^{2}\right) }+\frac{R_{1}M_{1}^{2}%
\left( M_{2}^{2}-M_{1}^{2}\right) }{\left( M_{1}^{2}-q^{2}\right) \left(
M_{2}^{2}-q^{2}\right) }+\frac{1}{M_{2}^{2}-q^{2}}\frac{1}{\pi }%
\int_{S_{0}}^{M^{2}}\frac{M_{2}^{2}-s}{s-q^{2}}{\Im} F_{V}^{{Cont}}
\label{26g}
\end{equation}
If we model the continuum contribution by quark triangle graph as given in
Eq. (\ref{26}), we obtain 
\begin{equation}
F\left( q^{2}\right) =\frac{RM^{2}\left( M_{2}^{2}-M^{2}\right) }{\left(
M^{2}-q^{2}\right) \left( M_{2}^{2}-q^{2}\right) }+\frac{R_{1}M_{1}^{2}%
\left( M_{2}^{2}-M_{1}^{2}\right) }{\left( M_{1}^{2}-q^{2}\right) \left(
M_{2}^{2}-q^{2}\right) }+\frac{M_{2}^{2}-M^{2}}{\left(
M_{2}^{2}-q^{2}\right) \left( M^{2}-q^{2}\right) }c  \label{26h}
\end{equation}
where in the heavy quark limit $M_{B}=M_{B}^{*}=M$ and 
\begin{equation}
c=f_{B}M_{B}\left[ \frac{Q_{u}}{\bar{\Lambda}}+{\cal O}\left( \frac{1}{M_{B}}%
\right) \right]  \label{26i}
\end{equation}

\section{Ward Identities Constraints}

It is useful to define 
\begin{eqnarray}
\left\langle \gamma \left( k,\epsilon \right) \left| \bar{u}i\sigma ^{\mu
\nu }q_{\nu }b\right| B(p)\right\rangle &=&-i\varepsilon ^{\mu \nu \alpha
\beta }\epsilon _{\nu }^{*}k_{\alpha }p_{\beta }F_{1}(q^{2})  \label{27} \\
\left\langle \gamma \left( k,\epsilon \right) \left| \bar{u}i\sigma ^{\mu
\nu }\gamma _{5}q_{\nu }b\right| B(p)\right\rangle &=&\left[ \left( q\cdot
k\right) \epsilon ^{*\mu }-\left( \epsilon ^{*}\cdot q\right) k^{\mu
}\right] F_{3}(q^{2})  \label{28}
\end{eqnarray}
Now we will make use of Ward Identities and gauge invariance principle to
relate different form factors.

Usually, the gauge invariance is implemented by means of the Ward
Identities; another way, essentially the same, is to consider what happens
if the polarization vector of an external (real) photon is replaced by its
four-momentum. The result is zero, provided that one considers all diagrams
where this particular photon is connected in all possible ways to a charge
carrying line. In this way one understands the connection between gauge
invariance and charge conservation. The Ward Identities\footnote{%
See ref.\cite{JHEP092003065} for a detailed derivation of these Ward
Identities.} used to relate different form factors appearing in our process
are: 
\begin{eqnarray}
\left\langle \gamma \left( k,\epsilon \right) \left| \bar{u}i\sigma ^{\mu
\nu }q_{\nu }b\right| B(p)\right\rangle &=&-(m_{b}+m_{q})\left\langle \gamma
\left( k,\epsilon \right) \left| \bar{u}\gamma ^{\mu }b\right|
B(p)\right\rangle  \nonumber \\
&&+(p^{\mu }+k^{\mu })\left\langle \gamma \left( k,\epsilon \right) \left| 
\bar{u}b\right| B(p)\right\rangle  \nonumber \\
&=&-(m_{b}+m_{q})\left\langle \gamma \left( k,\epsilon \right) \left| \bar{u}%
\gamma ^{\mu }b\right| B(p)\right\rangle  \label{29} \\
\left\langle \gamma \left( k,\epsilon \right) \left| \bar{u}i\sigma ^{\mu
\nu }\gamma _{5}q_{\nu }b\right| B(p)\right\rangle
&=&(m_{b}-m_{q})\left\langle \gamma \left( k,\epsilon \right) \left| \bar{u}%
\gamma ^{\mu }\gamma _{5}b\right| B(p)\right\rangle  \nonumber \\
&&+(p^{\mu }+k^{\mu })\left\langle \gamma \left( k,\epsilon \right) \left| 
\bar{u}\gamma _{5}b\right| B(p)\right\rangle  \nonumber \\
&=&(m_{b}-m_{q})\left\langle \gamma \left( k,\epsilon \right) \left| \bar{u}%
\gamma ^{\mu }\gamma _{5}b\right| B(p)\right\rangle  \label{30}
\end{eqnarray}
where the matrix elements $\left\langle \gamma \left( k,\epsilon \right)
\left| \bar{u}b\right| B(p)\right\rangle $ and $\left\langle \gamma \left(
k,\epsilon \right) \left| \bar{u}\gamma _{5}b\right| B(p)\right\rangle $
vanish for real photon due to gauge invariance.

Using the Ward Identities in Eqs.(\ref{27}) and (\ref{28}), and comparing
the coefficients, we obtain [$p\cdot k=q\cdot k$, $\epsilon ^{*}\cdot
p=\epsilon ^{*}\cdot q$] 
\begin{eqnarray}
F_{V}(q^{2}) &=&\frac{1}{m_{b}+m_{q}}F_{1}(q^{2})  \label{31} \\
F_{A}(q^{2}) &=&\frac{1}{m_{b}-m_{q}}F_{3}(q^{2})  \label{32}
\end{eqnarray}
The results given in Eqs.(\ref{31}) and (\ref{32}) are model independent
because these are derived by using Ward Identities.

In order to make use of Ward Identities to relate different form factors, we
define 
\begin{eqnarray}
\left\langle \gamma \left( k,\epsilon \right) \left| i\bar{u}\sigma _{\alpha
\beta }b\right| B(p)\right\rangle &=&-i\varepsilon _{\alpha \beta \rho
\sigma }\epsilon ^{*\rho }(k)\left[ (p+k)^{\sigma }g_{+}+q^{\sigma
}g_{-}\right]  \nonumber \\
&&-iq\cdot \epsilon ^{*}(k)\varepsilon _{\alpha \beta \rho \sigma
}(p+k)^{\rho }q^{\sigma }h  \nonumber \\
&&-i\left[ q_{\alpha }\varepsilon _{\beta \rho \sigma \tau }\epsilon ^{*\rho
}(k)(p+k)^{\sigma }q^{\tau }-\alpha \leftrightarrow \beta \right] h_{1} 
\nonumber \\
&&-i\left[ \left( p+k\right) _{\alpha }\varepsilon _{\beta \rho \sigma \tau
}\epsilon ^{*\rho }(k)\left( p+k\right) ^{\sigma }q^{\tau }-\alpha
\leftrightarrow \beta \right] h_{2}.  \nonumber \\
&&  \label{33}
\end{eqnarray}
Since we have a real photon, gauge invariance requires that if we replace $%
\epsilon ^{\mu }(k)$ by $k^{\mu }$, the matrix element should vanish. This
requires 
\begin{equation}
g_{+}+g_{-}+2\left( q\cdot k\right) h=0  \label{34}
\end{equation}
From Dirac algebra 
\begin{equation}
\sigma ^{\mu \nu }\gamma _{5}=-\frac{i}{2}\varepsilon ^{\mu \nu \alpha \beta
}\sigma _{\alpha \beta },  \label{35}
\end{equation}
we can write 
\begin{eqnarray}
&&\left\langle \gamma \left( k,\epsilon \right) \left| i\bar{u}\sigma ^{\mu
\nu }\gamma _{5}b\right| B(p)\right\rangle  \nonumber \\
&=&-\frac{i}{2}\varepsilon ^{\mu \nu \alpha \beta }\left\langle \gamma
(k,\epsilon )\left| i\bar{u}\sigma _{\alpha \beta }b\right| B(p)\right\rangle
\nonumber \\
&=&\left( \epsilon ^{*\mu }k^{\nu }-\epsilon ^{*\nu }k^{\mu }\right) \left[
g_{+}-g_{-}-\left( M_{B}^{2}+q^{2}\right) h_{1}-\left(
3M_{B}^{2}-q^{2}\right) h_{2}\right]  \nonumber \\
&&+\left( \epsilon ^{*\mu }p^{\nu }-\epsilon ^{*\nu }p^{\mu }\right) \left[
g_{+}+g_{-}+\left( M_{B}^{2}-q^{2}\right) \left( h_{1}+h_{2}\right) \right] 
\nonumber \\
&&-2q\cdot \epsilon ^{*}\left( h-h_{1}-h_{2}\right) \left( p^{\mu }k^{\nu
}-p^{\nu }k^{\mu }\right)  \label{36}
\end{eqnarray}
The gauge invariance, namely, replacing $\epsilon ^{\mu }$ by $k^{\mu }$,
the matrix element should be zero, does not give any new relation other than
(\ref{34}). Using this relation and $2k\cdot q=M_{B}^{2}-q^{2}$, we get 
\begin{eqnarray}
&&\left\langle \gamma \left( k,\epsilon \right) \left| i\bar{u}\sigma ^{\mu
\nu }\gamma _{5}b\right| B(p)\right\rangle  \nonumber \\
&=&\left( \epsilon ^{*\mu }k^{\nu }-\epsilon ^{*\nu }k^{\mu }\right) \left[
2g_{+}+\left( M_{B}^{2}-q^{2}\right) \left( h-h_{1}-h_{2}\right)
-2q^{2}h_{1}-2M_{B}^{2}h_{2}\right]  \nonumber \\
&&-\left[ 2k\cdot q\left( \epsilon ^{*\mu }p^{\nu }-\epsilon ^{*\nu }p^{\mu
}\right) +2q\cdot \epsilon ^{*}\left( p^{\mu }k^{\nu }-p^{\nu }k^{\mu
}\right) \right] \left( h-h_{1}-h_{2}\right)  \label{37}
\end{eqnarray}
Contrary to what is stated in some literature, the gauge invariance does
allow a second tensor structure in addition to $\left( \epsilon ^{\mu
}k^{\nu }-\epsilon ^{\nu }k^{\mu }\right) $.

This gives 
\begin{eqnarray}
\left\langle \gamma \left( k,\epsilon \right) \left| i\bar{q}\sigma ^{\mu
\nu }q_{\nu }\gamma _{5}b\right| B(p)\right\rangle &=&2\left(
g_{+}-q^{2}h-\left( M_{B}^{2}-q^{2}\right) h_{2}\right)  \nonumber \\
&&\times \left( q\cdot k\epsilon ^{*\mu }\left( k\right) -q\cdot \epsilon
^{*}\left( k\right) k^{\mu }\right) .  \label{38}
\end{eqnarray}
This, in turn, gives [from Eq.(\ref{28})] 
\begin{equation}
F_{3}(q^{2})=2\left[ -g_{+}-q^{2}h-\left( M_{B}^{2}-q^{2}\right) h_{2}\right]
\label{39}
\end{equation}
Similarly, from Eq.(\ref{33}), we get the relation 
\[
\left\langle \gamma \left( k,\epsilon \right) \left| \bar{u}i\sigma _{\alpha
\beta }q^{\beta }b\right| B(p)\right\rangle =-i\varepsilon _{\alpha \beta
\rho \sigma }\epsilon ^{*\rho }q^{\beta }p^{\sigma }2\left[
g_{+}-q^{2}h_{1}-M_{B}^{2}h_{2}\right] 
\]
Comparison of this equation with Eq.(\ref{27}) gives 
\begin{equation}
F_{1}\left( q^{2}\right) =2[g_{+}\left( q^{2}\right) -q^{2}h_{1}\left(
q^{2}\right) -M_{B}^{2}h_{2}\left( q^{2}\right) ]  \label{40}
\end{equation}
Thus, finally we obtain 
\begin{eqnarray}
F_{V}\left( q^{2}\right) &=&\frac{2}{m_{b}+m_{q}}\left\{ g_{+}\left(
q^{2}\right) -q^{2}h_{1}\left( q^{2}\right) -M_{B}^{2}h_{2}\left(
q^{2}\right) \right\} ,  \label{41} \\
F_{A}\left( q^{2}\right) &=&\frac{2}{m_{b}-m_{q}}\left\{ g_{+}\left(
q^{2}\right) -q^{2}h\left( q^{2}\right) -\left( M_{B}^{2}-q^{2}\right)
h_{2}\left( q^{2}\right) \right\} .  \label{42}
\end{eqnarray}
Therefore, the normalization of $F_{V}$ and $F_{A}$ at $q^{2}=0$ is
determined by a universal form factor $\left( g_{+}\left( 0\right)
-M_{B}^{2}h_{2}\right) $. Now the form factor $h_{2}$ does not get any
contribution from quark triangle graph nor from the pole and therefore we
shall put it equal to zero. On the other hand, only $g_{+}\left(
q^{2}\right) $ gets contribution from quark $\Delta $-graph \cite
{IJMPA17-4927}, 
\begin{equation}
g_{+}\left( q^{2}\right) =f_{B}\left\{ \frac{Q_{u}}{2\bar{\Lambda}}-\frac{%
Q_{b}}{2M_{B}}\left( 1-\frac{m_{q}}{M_{B}}\right) \right\} \frac{1}{%
1-q^{2}/M_{B}^{2}}.  \label{46}
\end{equation}
We expect the Ward Identities to hold at low $q^{2}$ below the resonance
regime and as such we use the results obtained from them at $q^{2}=0$. Thus
from Eqs. (\ref{41} and \ref{42}), we obtain 
\begin{equation}
\left( m_{b}+m_{q}\right) F_{V}\left( 0\right) =2g_{+}\left( 0\right)
=\left( m_{b}-m_{q}\right) F_{A}\left( 0\right) .  \label{47}
\end{equation}
Further, using Eq. (\ref{26Lambda}) in the above Eq. (\ref{47}) and
neglecting terms of the order of $\left( \bar{\Lambda}\mp m_{q}\right) $ $%
/M_{B}$, we obtain another constraint using Eqs. (\ref{26h}, \ref{26i}) at $%
q^{2}=0$ 
\begin{equation}
R\left( 1-\frac{M^{2}}{M_{2}^{2}}\right) +R_{1}\left( 1-\frac{M_{1}^{2}}{%
M_{2}^{2}}\right) =\left( \frac{2g_{+}\left( 0\right) }{M}\right) \frac{M^{2}%
}{M_{2}^{2}}.  \label{48}
\end{equation}
Now if we restrict to one radial excitation $\left( M_{2}=M_{1}\right) $ we
obtain from Eq. (\ref{48}) 
\begin{eqnarray}
R &=&\frac{2g_{+}\left( 0\right) }{\left( M_{1}^{2}/M^{2}-1\right) M}
\label{49} \\
F\left( q^{2}\right) &=&\frac{2}{M}\frac{g_{+}\left( 0\right) }{\left(
1-q^{2}/M^{2}\right) \left( 1-q^{2}/M_{1}^{2}\right) }  \label{50}
\end{eqnarray}
Restoring the subscripts and using the definitions (\ref{25}) 
\begin{eqnarray}
g_{B^{*}B\gamma } &=&\frac{2g_{+}(0)}{M_{B}}\frac{M_{B^{*}}^{2}}{%
f_{B^{*}}\left( M_{B_{1}^{*}}^{2}/M_{B^{*}}^{2}-1\right) }  \nonumber \\
&\simeq &\frac{2g_{+}(0)}{f_{B}\left(
M_{B_{1}^{*}}^{2}/M_{B^{*}}^{2}-1\right) }  \label{67}
\end{eqnarray}
while 
\begin{equation}
f_{B_{A}^{*}B\gamma }=\frac{M_{B_{A}^{*}}^{2}}{M_{B}}\frac{2g_{+}(0)}{%
f_{B_{A}^{*}}\left( M_{B_{A_{1}}^{*}}^{2}/M_{B_{A}^{*}}^{2}-1\right) }
\label{68}
\end{equation}
Use $g_{+}\left( 0\right) $ given in Eq. (\ref{46}) with $Q_{u}=2/3$, namely 
\begin{equation}
g_{+}\left( 0\right) =\frac{2}{3}\frac{f_{B}}{2\bar{\Lambda}}  \label{69}
\end{equation}
we have the prediction 
\begin{equation}
g_{B^{*}B\gamma }=\frac{2}{3\bar{\Lambda}}\frac{1}{\left(
M_{B_{1}^{*}}^{2}/M_{B^{*}}^{2}-1\right) }  \label{70}
\end{equation}
Further 
\begin{eqnarray}
F_{V}(q^{2}) &=&\frac{2}{M_{B}}\frac{g_{+}\left( 0\right) }{\left(
1-q^{2}/M_{B^{*}}^{2}\right) \left( 1-q^{2}/M_{B_{1}^{*}}^{2}\right) }
\label{71} \\
F_{A}(q^{2}) &=&\frac{2}{M_{B}}\frac{g_{+}\left( 0\right) }{\left(
1-q^{2}/M_{B_{A}^{*}}^{2}\right) \left( 1-q^{2}/M_{B_{A_{1}}^{*}}^{2}\right) 
}  \label{72}
\end{eqnarray}
This is the final expression for the form factors of our process $%
B\rightarrow \gamma l\nu _{l}$, if we restrict to the one radial excitation.
We also observe the approximate equality $F_{V}(q^{2})=F_{A}(q^{2})$ of the
form factors which also occur in some other models \cite
{PRD61-114510,PRD64-097503}. For numerical work, we shall use $B$-meson
masses given in Table \ref{table1} and $f_{B}=0.180$ GeV. 
\begin{table}[tbp]
\caption{$B$-mesons masses in GeV \protect\cite{PRD21-203} }
\label{table1}\vspace{0.1cm}
\par
\begin{center}
\begin{tabular}{c}
\begin{tabular}{|c|c|c|c|c|}
\hline
& $J^P$ & $M$ & $M_1/M$ & $M_2/M$ \\ \hline
$M_B$ & $0^{-}$ & $5.28$ & $1.14$ & 1.24 \\ \hline
$M_{B^{*}}$ & $1^{-}$ & $5.33$ & $1.14$ & 1.24 \\ \hline
$M_{B_A^{*}}$ & $1^{+}$ & $5.71$ & $1.12$ & 1.22 \\ \hline
\end{tabular}
\end{tabular}
\end{center}
\end{table}

This gives the prediction from Eq.(\ref{70}) 
\begin{equation}
g_{B^{*}B\gamma }=\frac{2.2}{\bar{\Lambda}}=5.6{ GeV}^{-1},  \label{73}
\end{equation}
for $\bar{\Lambda}=5.28-4.8=0.4$ GeV$^{-1}$ [see Eq. (\ref{26Lambda}) and
Table 1]. Also, we obtain from Eq. (\ref{69}) 
\begin{equation}
g_{+}(0)=\frac{3}{20}=0.15.  \label{74}
\end{equation}
Further from Eq.(\ref{68}) 
\begin{eqnarray}
f_{B_{A}^{*}B\gamma } &=&\frac{f_{B}M_{B_{A}^{*}}}{f_{B_{A}^{*}}}\frac{2.6}{%
\bar{\Lambda}}  \nonumber \\
&=&6.5\frac{f_{B}M_{B_{A}^{*}}}{f_{B_{A}^{*}}}{ GeV}^{-1}  \label{75}
\end{eqnarray}

We now study the effect of the second radial excitation. We go back to Eq. (%
\ref{26h}) and use the constraint (\ref{48}) to obtain 
\[
F\left( q^{2}\right) =\frac{R\left( \frac{M_{2}^{2}}{M^{2}}-1\right) \left( 
\frac{M_{1}^{2}}{M^{2}}-1\right) \frac{M^{2}}{M_{2}^{2}}\frac{q^{2}}{%
M_{1}^{2}}+\frac{2g_{+}\left( 0\right) }{M}\left( 1-q^{2}\left( \frac{1}{%
M_{2}^{2}}+\frac{1}{M_{1}^{2}}-\frac{M^{2}}{M_{1}^{2}M_{2}^{2}}\right)
\right) }{\left( 1-q^{2}/M_{2}^{2}\right) \left( 1-q^{2}/M_{1}^{2}\right)
\left( 1-q^{2}/M^{2}\right) } 
\]
If we parametrize $R$ as 
\[
R=\frac{2g_{+}\left( 0\right) }{M}\frac{1-\left(
1-M_{1}^{2}/M_{2}^{2}\right) A}{\left( M_{1}^{2}/M^{2}-1\right) }, 
\]
where $A$ is a parameter which in principle can be obtained when $%
g_{B^{*}B\gamma }$ and $f_{B_{A}^{*}B\gamma }$ become known. Then 
\begin{equation}
F\left( q^{2}\right) =\frac{2g_{+}\left( 0\right) }{M}\frac{1-\frac{q^{2}}{%
M_{1}^{2}}\left( 1+\left( 1-\frac{M^{2}}{M_{2}^{2}}\right) \left( 1-\frac{%
M_{1}^{2}}{M_{2}^{2}}\right) A\right) }{\left( 1-q^{2}/M_{2}^{2}\right)
\left( 1-q^{2}/M_{1}^{2}\right) \left( 1-q^{2}/M^{2}\right) }
\label{ff-Final}
\end{equation}
For $M_{1}=M_{2}$ the above Eq. (\ref{ff-Final}) reduces to Eq. (\ref{50}).
So the couplings of $B$ with $B^{*}\gamma $ and $B_{A}^{*}\gamma $ become 
\begin{eqnarray}
g_{B^{*}B\gamma } &=&\frac{2g_{+}(0)M_{B^{*}}^{2}}{M_{B}f_{B^{*}}\left(
M_{B_{1}^{*}}^{2}/M_{B^{*}}^{2}-1\right) }\left[ 1-\left(
1-M_{B^{*}}^{2}/M_{B_{1}^{*}}^{2}\right) \left(
1-M_{B_{1}^{*}}^{2}/M_{B^{*}}^{2}\right) A\right]  \nonumber \\
&=&\left[ 1-\left( 1-M_{B^{*}}^{2}/M_{B_{1}^{*}}^{2}\right) \left(
1-M_{B_{1}^{*}}^{2}/M_{B^{*}}^{2}\right) A\right] 5.6{ GeV}^{-1}
\label{ncoupling1} \\
f_{B_{A}^{*}B\gamma } &=&\frac{f_{B}M_{B_{A}^{*}}}{f_{B_{A}^{*}}}\left[
1-\left( 1-M_{B_{A}^{*}}^{2}/M_{B_{A_{1}}^{*}}^{2}\right) \left(
1-M_{B_{A_{1}}^{*}}^{2}/M_{B_{A}^{*}}^{2}\right) A\right] 6.5{ GeV}^{-1}
\nonumber \\
&&  \label{ncoupling2}
\end{eqnarray}
and the corresponding form factors become 
\begin{eqnarray}
F_{V}(q^{2}) &=&\frac{2g_{+}\left( 0\right) }{M_{B}}\frac{1-\frac{q^{2}}{%
M_{B_{1}^{*}}^{2}}\left( 1+\left( 1-\frac{M_{B^{*}}^{2}}{M_{B_{2}^{*}}^{2}}%
\right) \left( 1-\frac{M_{B_{1}^{*}}^{2}}{M_{B_{2}^{*}}^{2}}\right) A\right) 
}{\left( 1-q^{2}/M_{B_{2}^{*}}^{2}\right) \left(
1-q^{2}/M_{B_{1}^{*}}^{2}\right) \left( 1-q^{2}/M_{B^{*}}^{2}\right) }
\label{formfactor1} \\
F_{A}(q^{2}) &=&\frac{2g_{+}\left( 0\right) }{M_{B}}\frac{1-\frac{q^{2}}{%
M_{B_{A_{1}}^{*}}^{2}}\left( 1+\left( 1-\frac{M_{B_{A}^{*}}^{2}}{%
M_{B_{A_{2}}^{*}}^{2}}\right) \left( 1-\frac{M_{B_{A_{1}}^{*}}^{2}}{%
M_{B_{A_{2}}^{*}}^{2}}\right) A\right) }{\left(
1-q^{2}/M_{B_{A_{2}}^{*}}^{2}\right) \left(
1-q^{2}/M_{B_{A_{1}}^{*}}^{2}\right) \left( 1-q^{2}/M_{B_{A}^{*}}^{2}\right) 
}  \label{formfactor2}
\end{eqnarray}
For numerical values we shall use $A=0$ $[$i.e., $M_{1}=M_{2}]$ and $A=3$
and $A=4.8$. The second value of $A\left( =3\right) $ corresponds to
estimate of $g_{B^{*}B\gamma }$ from vector meson dominance 
\[
g_{B^{*}B\gamma }=\frac{2}{3}g_{B^{*}B\rho ^{-}}\frac{f_{\rho ^{-}}}{m_{\rho
}^{2}}=2.76{ GeV}^{-1} 
\]
where $g_{B^{*}B\rho ^{-}}=\sqrt{2}(11)$ GeV$^{-1}$ obtained in \cite
{Dominguez} and $f_{\rho ^{-}}/m_{\rho }=205$ MeV. The third value of $%
A\left( =4.8\right) $ gives more or less the width for $B^{*}\rightarrow
B\gamma $ obtained from MI transition in non relativistic quark model
(NRQM). These values give decay width for $B^{*}\rightarrow B\gamma $
transition $23$ keV, 5.5 keV and $0.8$ keV respectively while MI transition
in NRQM predicts it to be $0.9$ keV. These predictions are testable when
above decay width is experimentally measured.

\section{Decay distribution}

The Dalitz plot density 
\begin{eqnarray}
\rho (x,y) &=&\frac{d^{2}\Gamma }{dxdy}=\frac{d^{2}\Gamma _{IB}}{dxdy}+\frac{%
d^{2}\Gamma _{SD}}{dxdy}+\frac{d^{2}\Gamma _{INT}}{dxdy}  \nonumber \\
&=&\rho _{IB}(x,y)+\rho _{SD}(x,y)+\rho _{INT}(x,y)  \label{3.2.8}
\end{eqnarray}
is a Lorentz invariant which contains the form factors $F_{V}$ and $F_{A}$
in the following form \cite{PRB136-1160,PRs88-151,NPB396-81} 
\begin{eqnarray*}
\rho _{IB}(x,y) &=&A_{IB}f_{IB}(x,y) \\
\rho _{SD}(x,y) &=&A_{SD}M_{B}^{2}\left[
(F_{V}+F_{A})^{2}f_{SD^{+}}(x,y)+(F_{V}-F_{A})^{2}f_{SD^{-}}(x,y)\right] \\
\rho _{INT}(x,y) &=&A_{INT}M_{B}\left[
(F_{V}+F_{A})f_{INT^{+}}(x,y)+(F_{V}-F_{A})f_{INT^{-}}(x,y)\right]
\end{eqnarray*}
where 
\begin{eqnarray*}
f_{IB}(x,y) &=&\left( \frac{1-y+r_{l}}{x^{2}(x+y-1-r_{l})}\right) \\
&&\times \left( x^{2}+2\left( 1-x\right) \left( 1-r_{l}\right) -\frac{%
2xr_{l}\left( 1-r_{l}\right) }{(x+y-1-r_{l})}\right) \\
f_{SD^{+}}(x,y) &=&(x+y-1-r_{l})\left( (x+y-1)\left( 1-x\right) -r_{l}\right)
\\
f_{SD^{-}}(x,y) &=&(1-y+r_{l})\left( \left( 1-x\right) (1-y)+r_{l}\right) \\
f_{INT^{+}}(x,y) &=&\left( \frac{1-y+r_{l}}{x(x+y-1-r_{l})}\right) \left(
\left( 1-x\right) (1-x-y)+r_{l}\right) \\
f_{INT^{-}}(x,y) &=&\left( \frac{1-y+r_{l}}{x(x+y-1-r_{l})}\right) \left(
x^{2}-\left( 1-x\right) (1-x-y)-r_{l}\right)
\end{eqnarray*}
and 
\begin{eqnarray*}
{A}_{IB} &=&4r_{l}\left( \frac{f_{B}}{M_{B}}\right) ^{2}{A}_{SD} \\
{A}_{SD} &=&\frac{G_{F}^{2}}{2}\frac{\left| {V}_{ub}\right| ^{2}\alpha }{%
32\pi ^{2}}M_{B}^{5} \\
{A}_{INT} &=&4r_{l}\left( \frac{f_{B}}{M_{B}}\right) {A}_{SD}
\end{eqnarray*}

The $SD^{+}$ term reaches its maximum at $x=2/3,$ $y=1$, which corresponds
to $\theta _{l\gamma }=\pi $. The $SD^{-}$ term reaches its maximum at $%
x=2/3 $, $y=1/3$, corresponding to $\theta _{l\gamma }=0$. Indeed, for
lepton of maximal energy ($y=1$), only ``right-handed'' photons contribute.
In this situation, the photon and the neutrino must be emitted in the
direction opposite to that of the lepton. Angular momentum conservation
forces the photon spin to be opposite to the total lepton spin and the
photon helicity has the same sign as that of the lepton. Then the photon and
the neutrino are emitted parallel. This configuration corresponds to a
neutrino of maximal energy ($E_{\nu }=E_{\nu }^{\max }$ when $x+y=1$). In
this case, only the ``left-handed'' photon contributes. When $x+y=1$, the $%
IB $ contribution becomes very large: this corresponds to $\theta _{l\gamma
}=0$. Consequently, it is very difficult to distinguish experimentally
between the $IB$ and the $SD^{-}$ contribution. To summarize, an experiment
performed in the region $\theta _{l\gamma }\simeq \pi $ is essentially
sensitive to $(F_{V}+F_{A})^{2}$.

The form factors calculated in Eq. (\ref{50}) can be expressed in terms of
dimensionless variable $x$, 
\begin{equation}
F(x)=\frac{F\left( 0\right) }{x\left[ 1-\left( 1-x\right) /\left(
M_{1}/M\right) ^{2}\right] },  \label{4.5.1}
\end{equation}
where $x$ is defined in Eq. (\ref{12}) and $q^{2}$ in Eq. (\ref{15}). After
restoring subscripts, the form factors $F_{V}\left( q^{2}\right) $ [Eq. (\ref
{71})] and $F_{A}\left( q^{2}\right) $[Eq. (\ref{72})] can be written as 
\begin{eqnarray}
F_{V}(x) &=&\frac{F_{V}\left( 0\right) }{x\left[ 1-\left( 1-x\right) /\left(
M_{B_{1}^{*}}/M_{B^{*}}\right) ^{2}\right] }  \label{decayffactor1} \\
F_{A}(x) &=&\frac{F_{V}\left( 0\right) }{x\left[ 1-\left( 1-x\right) /\left(
M_{B_{A_{1}}^{*}}/M_{B_{A}^{*}}\right) ^{2}\right] },  \label{decayffactor2}
\end{eqnarray}
where 
\[
F_{V,A}\left( 0\right) =\frac{2g_{+}\left( 0\right) }{M_{B}} 
\]
We use these in Eq. (\ref{3.2.8}) and integrate over $x$ and $y$ in the
limit as mentioned in Eqs. (\ref{16}, \ref{17}). IB contribution diverges
for the minimum value of $x$, we take an arbitrary lower limit for $x$ i.e. $%
x_{\min }\approx r_{l}$ for which the divergence problem is cured and the IB
part gives some definite value ${\cal O}(10^{-20})$. But as the energy of
the photon is increased, it approaches zero at $x_{\max }$. Therefore in the
total decay width, this does not contribute much. The $SD$ part is the most
dominant part of the decay width which provides almost the whole
contribution. This part increases initially with increasing $x$, reaches its
peak value and then starts decreasing. The $INT$ part of the decay width is
an increasingly vanishing contribution and can be neglected in comparison to
the $SD$\ part, because it is suppressed by ${\cal O}(10^{-21})$ and becomes
flat (approaches zero) as $x$ (the photon energy) approaches $1$ (its
maxima). Therefore, this does not contribute fairly to the total decay width
of the process.

In the Fig. \ref{compar1}, differential decay width of the process is
plotted against $x$ and we see that for our calculations, the peak is
shifted to lower value of $x$ as compared to those for Eilam et al., \cite
{PLB361-137}, Korchemsky et al. \cite{PRD61-114510} and Chelkov et al. \cite
{PRD64-097503}. So, for the process $B\rightarrow \gamma l\nu _{l}$ the
branching ratios obtained is 
\begin{equation}
{\cal B}(B\rightarrow \gamma l\nu _{l})=
\begin{tabular}{ccc}
$0.5\times 10^{-6}$ &  & $(l=\mu )$%
\end{tabular}
\label{4.5.2}
\end{equation}
This value is for the form factors given in Eqs. (\ref{decayffactor1}, \ref
{decayffactor2}) which are obtained by restricting to the first radial
excitation only. Now if we consider the effect of second radial excitation
the expression for the form factors are given in Eqs. (\ref{formfactor1}, 
\ref{formfactor2}). The branching ratio thus obtained are 
\begin{eqnarray*}
{\cal B}(B &\rightarrow &\gamma l\nu _{l})=
\begin{tabular}{ccc}
$0.38\times 10^{-6}$ &  & $(l=\mu ,\,A=3.0)$%
\end{tabular}
\\
{\cal B}(B &\rightarrow &\gamma l\nu _{l})=
\begin{tabular}{ccc}
$0.32\times 10^{-6}$ &  & $(l=\mu ,\,A=4.8)$%
\end{tabular}
\end{eqnarray*}
for two representative cases of $A=3$ and $A=4.8$ respectively. These are
not sensitive to the values of $A$ in contrast to the decay width of $%
B^{*}\rightarrow B\gamma $. The CLEO\ collaboration indicate an upper limit
on the branching ratio ${\cal B}(B^{+}\rightarrow \gamma \nu _{l}e^{+})$ of $%
2.0\times 10^{-4}$ at the $90\%$ confidence level \cite{PRD56-11}. The
predicted values are within the upper limit provided by CLEO collaboration
but differ from those predicted in \cite{PRD61-114510,PRD64-097503}, namely $%
\left( 2-5\right) \times 10^{-6}$ and $0.9\times 10^{-6}$, respectively. The
Monte-Carlo simulation results are given in \cite{Lattery1996} where the
upper limit on the branching ratio for this process is predicted to be $%
5.2\times 10^{-5}$.

\section{Conclusions}

Preliminary data from the CLEO Collaboration indicate an upper limit on the
branching ratio ${\cal B}(B^{+}\rightarrow \gamma \nu _{l}e^{+})$ of $%
2.0\times 10^{-4}$ at the $90\%$ confidence level \cite{PRD56-11}. With the
better statistics expected from the upcoming $B$ factories, the observation
and experimental study of this decay could become soon feasible. It is
therefore of some interest to have a good theoretical control over the
theoretical uncertainties affecting the relevant matrix elements.

We have studied $B\rightarrow \gamma l\nu _{l}$ decay using dispersion
relations, asymptotic behavior of form factors and Ward Identities. The
dispersion relation involves ground state $B^{*}$ and $B_{A}^{*}$ resonances
and their radial excitations which model contributions from higher states
and continuum contribution, which is calculated from quark triangle graph.
The asymptotic behavior of form factors and Ward Identities fix the
normalization of the form factors in terms of universal function $%
g_{+}\left( 0\right) $ at $q^{2}=0$ and put constraints on the residues.
Thus in our approach, a parameterization of $q^{2}$ dependence of form
factors is not approximated by single pole contributions. This
parameterization is dictated by considerations mentioned above and also the
coupling constants of $1^{-}$($B^{*}$) and $1^{+}$($B_{A}^{*}$) resonances
with photon are predicted if we restrict to one radial excitation. By using $%
\bar{\Lambda}=0.4$ GeV$^{-1}$ we have calculated $g_{+}(0)=0.15$ and
predicted the value of $g_{B^{*}B\gamma }=5.6$ GeV$^{-1}$ (cf. Eq. (\ref{73}%
)) and $f_{B_{A}^{*}B\gamma }=6.5f_{B}M_{B_{A}^{*}}/f_{B_{A}^{*}}$ GeV$^{-1}$%
(cf. Eq. (\ref{75})). Taking into account one radial excitation the form
factors are summarized in Eq. (\ref{71}, \ref{72}). Branching ratio for the
process is then calculated to be ${\cal B}(B\rightarrow \gamma \nu
_{l}l)=0.5\times 10^{-6}$ which lies within the upper limit predicted by
CLEO Collaboration at $90\%$ confidence level \cite{PRD56-11}. Then we study
the effect of second radial excitation in terms of a single parameter $A$,
which in principle is determined once $g_{B^{*}B\gamma }$ and $%
f_{B_{A}^{*}B\gamma }$ are known (cf. Eq. (\ref{ncoupling1}, \ref{ncoupling2}%
)). The resulting form factors are given in Eqs. (\ref{formfactor1}, \ref
{formfactor2}). By using these form factors the branching ratio is ${\cal B}%
(B\rightarrow \gamma \nu _{l}l)=0.38\times 10^{-6}$ and ${\cal B}%
(B\rightarrow \gamma \nu _{l}l)=0.32\times 10^{-6}$ for two representative
cases $A=3.0$ and $A=4.8$ respectively. These branching ratios are not
sensitive to the value of $A$ in contrast to radiative coupling constants
which give respectively $B^{*}\rightarrow B\gamma $ width as $23$ keV ($A=0$%
), $5$ keV ($A=3.0$) and $0.8$ keV ($A=4.8$). One can also predict radiative
widths of radial excitation in terms of $B^{*}$ and $B_{A}^{*}$ radiative
widths by using relations (\ref{convergence2}), (\ref{26i}) and (\ref{48}).
The differential decay width versus photon energy is plotted in Fig. \ref
{compar1} to compare our results with the existing calculations in the
light-cone QCD approach \cite{PLB361-137,PRD61-114510} and in the
instantaneous Bethe-Salpeter approach \cite{PRD64-097503}. The results for $%
B\rightarrow \gamma \nu _{l}l$ have been reproduced by using Suddakov
resummation \cite{PRD61-114510} and have also been shown graphically. In our
calculations as well as in \cite{PLB361-137}, the position of the peak of
differential decay width is shifted towards the lower value of photon energy
spectrum. This is due to the double pole in the form factors. The over all
effect of radial excitations is to soften the $q^{2}$-behavior of
differential decay distribution while in \cite{PRD61-114510} it is due to
Suddakov resummation.

Our main inputs have been dispersion relations, asymptotic behavior and Ward
Identities, all of which have strong theoretical basis and in these aspects
it differs from other approaches. Our approach is closer to the one followed
in \cite{Burdman2} for $B\rightarrow \pi l\nu _{l}$. Only external
parameters involved are $f_{B}$, resonance masses (which are determined in
potential models) and $g_{B^{*}B\gamma }$ and $f_{B_{A}^{*}B\gamma }$ which
are either predicted or on which we have some theoretical information. The
radiative widths of radial excitations are predicted in terms of the above
coupling constants. Thus our approach has predictive power and can be tested
by future experiments.

The experiments at the B-factories, BaBar at SLAC and Belle at KEK (Japan)
and the planned hadronic accelerators are capable to measure the branching
ratio as low as $10^{-8}$ \cite{IC-2000-71}.

{\bf Acknowledgments:} The work of K, G and R was partially supported by
Pakistan Council of Science and Technology and that of A by ICSC-World
Laboratory fellowship.\ Authors K, A, and G thank Professor Fayyazuddin for
his valuable guidance and helpful discussions. R and A also acknowledge the
grant by Higher Education Commission (HEC) under National Distinguished
Professorship of R.

{\bf Figure Captions:}

\begin{enumerate}
\item  $B\rightarrow l\nu _{l}\gamma $ radiative leptonic decay diagrams. 
\label{rld1}

\item  Differential decay rate versus photon energy $x$ is plotted and
comparison is given with various approaches. The solid line (for $A=0$),
dashed-trippledotted line (for $A=3.0$) and dotted line (for $A=4.8$) are
our calculation, dash-dot-dot line \cite{PLB361-137}, dashed line \cite
{PRD61-114510} and dash-dotted line \cite{PRD64-097503}. \label{compar1}The
thin-solid line is the Sudekov resummation calculation result from Ref. \cite
{PRD61-114510}.
\end{enumerate}


\begin{thebibliography}{99}
\bibitem{PRD56-11}  CLEO\ Collaboration, T. E. Browder et al., Search for $%
B\rightarrow \mu \bar{\nu}_{\mu }\gamma $ and $B\rightarrow e\bar{\nu}%
_{e}\gamma $, Phys. Rev. D 56 (1997) 11; W. M. Yao et al., Journal of
Physics G 33 (2006) 1 

\bibitem{PRD51-111}  G. Burdman, T. Goldman, and D. Wyler, Radiative
leptonic decays of heavy mesons, Phys. Rev. D 51, 111 (1995)

\bibitem{PLB358-329}  A. Khodjamirian, G. Stoll, and D. Wyler, Calculation
of long distance effects in exclusive weak radiative decays of $B$ meson,
Phys. Lett. B 358, 129 (1995) [hep-ph/9506242]

\bibitem{NPB649-349}  E. Lunghi, D. Pirjol, and D. Wyler, Factorization in
leptonic radiative $B\rightarrow \gamma e\nu $ decays, Nucl. Phys. B 649
(2003) 349 [hep-ph/0210091]

\bibitem{NPB650-356}  S. Descotes-Genon, and C. T. Sachrajda, Factorization,
the light-cone distribution amplitude of the $B$-meson and the radiative
decay $B\rightarrow \gamma l\nu $, Nucl. Phys. B 650 (2003) 356
[hep-ph/0209216]

\bibitem{PLB372-331}  P. Colangelo, F. De Fazio, and G. Nardulli, On the
decay mode $B^{-}\rightarrow \mu ^{-}\bar{\nu}_{\mu }\gamma $, Phys. Lett. B
372, 331 (1996)

\bibitem{PLB386-328}  P. Colangelo, F. De Fazio, and G. Nardulli, Leptonic
constant from $B$ meson radiative decay, Phys. Lett. B 386, 328 (1996)
[hep-ph/9506332]

\bibitem{MPLA11-1061}  D. Atwood, G. Eilam, and A. Soni, Pure leptonic
radiative decays $B^{\pm }$, $D_{s}\rightarrow l\nu \gamma $ and the
annihilation graph, Mod. Phys. Lett. A 11, 1061 (1996)

\bibitem{JHEP042003029}  P. Ball, and E. Kou, $B\rightarrow \gamma e\nu $
transitions from QCD sum rules on the light-cone, JHEP 0304 (2003) 029

\bibitem{PLB361-137}  G. Eilam, I. Halperin, and R. R. Mendel, Radiative
decay $B\rightarrow l\nu \gamma $ in the light cone QCD approach, Phys.
Lett. B 361, 137 (1995)

\bibitem{PRD57-5697}  C. Q. Geng, C. C. Lih, and Wei-Min Zhang, Radiative
leptonic $B$ decays in the light front model, Phys. Rev. D 57, 5697 (1998)

\bibitem{PRD61-114510}  G. P. Korchemsky, D. Pirjol, and T.-M. Yan,
Radiative leptonic decays of $B$ mesons in QCD, Phys. Rev. D 61, 114510
(2000) [hep-ph/9911427]

\bibitem{PRD64-097503}  G. A. Chelkov, M. I. Gostkin and Z. K. Silagadze,
Radiative leptonic $B$ decays in the instantaneous Bethe-Salpeter approach,
Phys. Rev. D64, 097503 (2001)

\bibitem{PRL10-531}  N. Cabibbo, Unitary symmetry and leptonic decays, Phys.
Rev. Lett. 10, 531 (1963)

\bibitem{PTP49-652}  M. Kobayashi and T. Maskawa, CP-violation in the
renormalizable theory of weak interaction, Prog. Theor. Phys. 49, 652 (1973)

\bibitem{PRD15-709}  T. Goldmann, and W. J. Wilson, Radiative corrections to
leptonic decays of charged pseudoscalar mesons, Phys. Rev. D 15, 709 (1977)
and references therein.

\bibitem{SS9709500}  S. Stone, The goals and techniques of BTeV and LHC-B,
To appear in proceedings of ``Heavy Flavor Physics: A Probe of Nature's
Grand Design,'' Varenna, Italy, July 1997 [hep-ph/9709500]

\bibitem{Burdman2}  G. Burdman and J. Kambor, Dispersive approach to
semileptonic form factors in heavy-to-light meson decays, Phys. Rev. D 55
(1997) 2817

\bibitem{PRB136-1160}  S. G. Brown, and S. A. Bludman, Further analysis of
the decay $\pi \rightarrow e\nu \gamma $, Phys. Rev. B 136, 1160 (1964)

\bibitem{PRs88-151}  D. A. Bryman, P. Depommier, and C. Leroy, $\pi
\rightarrow e\nu $, $\pi \rightarrow e\nu \gamma $ decays and related
processes, Phys. Rep. 88, 151 (1982)

\bibitem{SJPN7-286}  D. Yu. Bardin and E. A. Ivanov, Weak-electromagnetic
decays $\pi \left( K\right) \rightarrow l\nu _{l}\gamma $ and $\pi \left(
K\right) \rightarrow l\nu _{l}l^{\prime +}l^{\prime -}$, Sov. J. Part. Nucl.
7, 286 (1976)

\bibitem{NPB396-81}  J. Bijnens, G. Ecker, and J. Gasser, Radiative
semileptonic kaon decays, Nucl. Phys. B 396, 81 (1993)

\bibitem{23}  Riazuddin, Role of gauge invariance in $B\rightarrow V\gamma $
radiative weak decays, Europhys. Lett. 60 (2002) 28

\bibitem{IJMPA17-4927}  See for details of calculations: Riazuddin, T.A.
Al-Aithan and A.H.S. Gilani, Form factors for $B\rightarrow \pi l\nu $ decay
in a model constrained by chiral symmetry and quark model, Int. J. Mod.
Phys. A 17 (2002) 4927 [hep-ph/0007164]; N. Paver and Riazuddin, The $\sigma 
$ and $\rho $ in $D$ and $B$ decays, [hep-ph/0107330]

\bibitem{nisgur}  N. Isgur, Phys. Rev. 13, 129 (1976); D23, 817E (1981)

\bibitem{Brodsky}  G. P. Lepage adn S. J. Brodsky, Phys. Lett. B 87 (1979)
359 and Phys. Rev. D 22 (1980) 2157

\bibitem{Henley}  A. Szcizepanick, E. M. Henley and S. J. Brodsky, Phys.
Lett. B 243 (1990) 287; G.Burdman and J. F. Donoghue, Phys. Lett. B 270
(1970) 55.

\bibitem{Quigg}  E. Eichten, C. T. Hill, and C. Quigg, in The Future of High
Energy Sensitive Charm Experiments, Proceeding of the Workshop, Batavia,
Illinois, 1994, edited by D. Kaplan and S. Kwan (Fermilab Report No. 94/190,
Batavia, 1994)

\bibitem{JHEP092003065}  A. H. S. Gilani, Riazuddin, and T. A. Al-Aithan,
Ward identities, $B\rightarrow \rho $ form factors and $\left| V_{ub}\right| 
$, JHEP 0309 (2003) 065 [hep-ph/0304183]

\bibitem{PRD21-203}  E. Eichten, K. Gottfried, T. Kinoshita, K.D. Lane and
T.-M. Yan, Charmonium: comparison with experiment, Phys. Rev. D 21 (1980)
203; M. Wirbel, B. Stech and M. Bauer, Exclusive semileptonic decays of
heavy mesons, Z. Physik C 29 (1985) 637.

\bibitem{Dominguez}  C. A. Dominguez and N. Paver, $QCD$ calculation of $%
b\rightarrow \pi $ lepton anti-lepton neutrino and the matrix element $V_{bu}
$, DESY Report No: DESY-88-063 (1988) and Z. Physik C 41 (1988) 217.

\bibitem{Lattery1996}  M. J. Lattery, Fully Leptonic Decays of $B$-Mesons,
Ph.D. thesis, University of Minnesota (1996); M. A. Shifman, Ups. Fiz. Nauk
151 (1987) 193 [Sov. Phys. Usp. 30 (1987) 91]

\bibitem{IC-2000-71}  G. G. Devidze, The short distance contribution to the $%
B_{s}\rightarrow \gamma \gamma $ decay in the SM and MSSM, ICTP preprint:
IC/2000/71 (2000); The lowest order short-distance contribution to the $%
B_{s}\rightarrow \gamma \gamma $, [hep-ph/9905431]
\end{thebibliography}
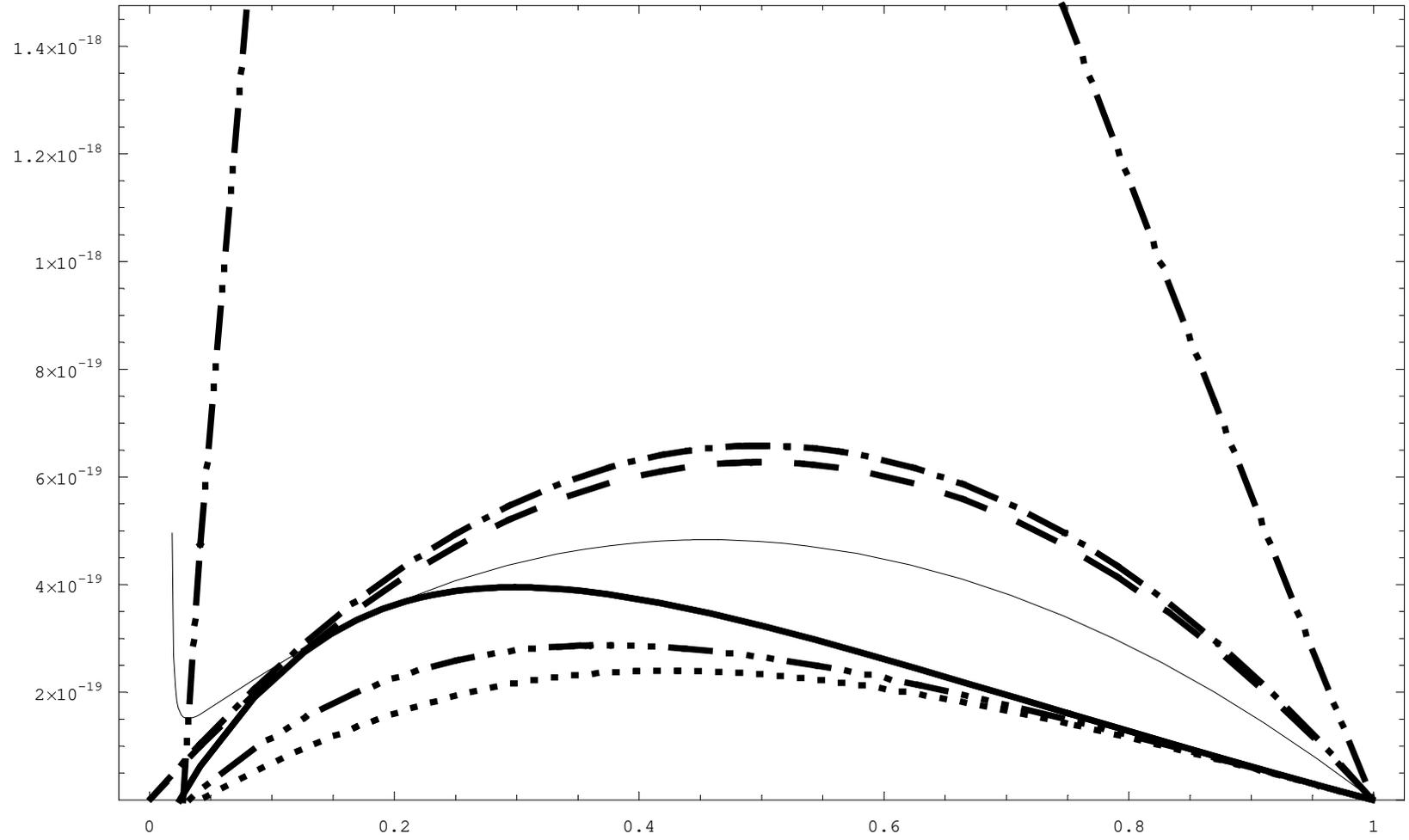
\end{document}